\documentclass[backend=biber]{gistt}

\usepackage{paralist}
\usepackage{listings}
\usepackage{todonotes}
\usepackage{xspace}
\usepackage{wrapfig}

\usetikzlibrary{calc, arrows, fit, shapes.geometric}

\definecolor{KiekerBlue}{RGB}{36,86,161}

\title{Interoperability From OpenTelemetry to Kieker:\\ Demonstrated as Export from the Astronomy Shop}
\author{David Georg Reichelt\\
Lancaster University Leipzig / \\
URZ Leipzig
\and
Shinhyung Yang\\
Kiel University\\
\and
Wilhelm Hasselbring\\
Kiel University
}

\begin{document}

\maketitle

\begin{abstract}
The observability framework Kieker provides a range of analysis capabilities, but it is currently only able to instrument a smaller selection of languages and technologies, including Java, C, Fortran, and Python. The OpenTelemetry standard aims for providing reference implementations for most programming languages, including C\# and JavaScript, that are currently not supported by Kieker. 
In this work, we describe how to transform OpenTelemetry tracing data into the Kieker framework. Thereby, it becomes possible to create for example call trees from OpenTelemetry instrumentations. We demonstrate the usability of our approach by visualizing trace data of the Astronomy Shop, which is an OpenTelemetry demo application.
\end{abstract}

\section{Introduction}

To understand the behavior of a software system, observability tools are used. Different observability tools provide different capabilities: While Kieker \cite{yang2025kieker} has various analysis capabilities and is known for its low overhead \cite{reichelt2021overhead}, OpenTelemetry is the de-facto standard for obtaining data and standard implementations provide agents for a variety of languages \cite{Blanco2023}. To make them interoperable, three steps are necessary:
\begin{inparaenum}[(1)]
  \item Transformation of Kieker traces into the OpenTelemetry format,
  \item Transformation of OpenTelemetry traces into the Kieker format, and
  \item Usage of native OpenTelemetry data formats in Kieker.
\end{inparaenum}
The first step has been done in our prior work \cite{reichelt2025kiekerToOtel}. 

In this work, we present the implementation of the second step. By transformation of OpenTelemetry traces into Kieker,\footnote{Traces are recevied via gRPC, e.g., from an agent.} we make it possible to use OpenTelemetries rich agent landscape within the Kieker analysis framework. Thereby, it becomes also feasible to do a range of Kieker analysis. Our analysis shows a structural difference between Kieker and OpenTelemetry: While Kieker traces are synchronous traces, supporting the observation of one control flow through a program, OpenTelemetry aims at representing asynchronous traces, supporting the observation of microservice calls. This results in different data structures and therefore limited compatibility of the concepts.

The remainder of the paper is structured as follows: First, we introduce the data formats of Kieker and OpenTelemetry. Based on this, we describe possible approaches for conversion of the data formats. Subsequently, we demonstrate the application of the conversion within the OpenTelemetry demo. Afterwards, we compare our approach to related work. Finally, we give a summary and an outlook.

\section{Data Formats}

In this section, we describe Kieker's and OpenTelemetry's data formats.

\paragraph{Kieker}

Within Kieker's monitoring part, the so-called monitoring log\footnote{In OpenTelemetry's terminology, this contains both tracing and monitoring information.} is obtained. This monitoring log is stored somewhere persistently to allow further analysis. These analyses consist of a variety of stages, that first read the monitoring data, transform them into traces and then provide analysis results, e.g., call trees and component graphs.

This architecture relies heavily on the data formats allowed within the monitoring log. These data formats are created using the Instrumentation Record Languages (IRL).\footnote{\url{https://github.com/kieker-monitoring/instrumentation-languages/wiki}} Based on an Xtext definition of records, the IRL allows to create source code for the usage of these records within the Kieker monitoring and analysis components.

\paragraph{OpenTelemetry}

OpenTelemetry provides data format standards for the three pillars of observability: Logs, metrics, and traces. Logs are timestamped text records, metrics are quantitive measurements of a system, and traces are a sequence of events which make it possible to follow an execution path. All OpenTelemetry data are serialized with protobuf serialization, ensuring high compression and low overhead for serialization and deserialization.\footnote{\url{https://github.com/open-telemetry/opentelemetry-proto}}

For observability, we consider \textbf{traces} the most important pillar. A trace consists of spans, which itself contains a name, a start and an end timestamp, a list of fields, a list of attributes and a list of events. The attributes are arbitrary key/value mappings, e.g., \lstinline'net.sock.peer.addr' might be mapped to \lstinline'127.0.0.1'. By the trace context level W3C recommendation,\footnote{\url{https://www.w3.org/TR/trace-context-2/}} that is implemented by OpenTelemetry, span contexts can be obtained across services.

Besides standard information like start and end time, spans can also contain attributes. These attributes are key-value pairs. The OpenTelemetry semantic conventions provide guidelines on how these attributes should be used.\footnote{\url{https://opentelemetry.io/docs/specs/semconv/general/attributes/}}

\section{Data Conversion}

For the basic data, most OpenTelemetry fields can just be mapped to their counterparts in Kieker. For the control flow, a more sophisticated approach is necessary.

\paragraph{Mappings}

\begin{figure}
  \begin{tikzpicture}[>=stealth, node distance=0.1cm, every node/.style={rectangle, draw, align=center, minimum width=3.3cm, minimum height=0.4cm, fill=white}]
    \node (tin) {tin};
    \node[above=0.5cm of tin.north, anchor=north, minimum width=3.9cm, rounded corners, fill=gray!20] {\textbf{Kieker}\\[3.4cm]};
    \node (tin) {tin};
    \node[below=of tin] (tout) {tout};
    \node[below=of tout] (ess) {ess};
    \node[below=of ess] (eoi) {eoi};
    \node[below=of eoi] (signature) {signature};
    \node[below=of signature] (hostname) {hostname};
    
    \node[left=1cm of tin] (startEpochNanos) {startEpochNanos};
    \node[above=0.5cm of startEpochNanos.north, anchor=north, minimum width=3.9cm, rounded corners, fill=gray!20] {\textbf{OpenTelemetry}\\[3.4cm]};
    \node[left=1cm of tin] (startEpochNanos) {startEpochNanos};
    \node[below=of startEpochNanos] (endEpochNanos) {endEpochNanos};
    \node[below=of endEpochNanos] (parent) {parent};
    \node[below=of parent] (name) {name};
    \node[below=of name] (serviceName) {net.sock.peer.addr};
    \node[below=of serviceName] (serviceId) {net.peer.name};
    
    \draw[<-] (tin)   -- (startEpochNanos);
    \draw[<-] (tout)   -- (endEpochNanos);
    \draw[<-] (eoi)   -- (parent);
    \draw[<-] (ess)   -- (parent);
    \draw[<-] (hostname)   -- (serviceName);
    \draw[<-] (hostname)   -- (serviceId);
    \draw[<-] (signature)   -- (name);
  \end{tikzpicture}
  \vspace{-0.75cm}
  \caption{Mapping of Fields}
  \label{fig:mapping}
  \vspace{-0.6cm}
\end{figure}
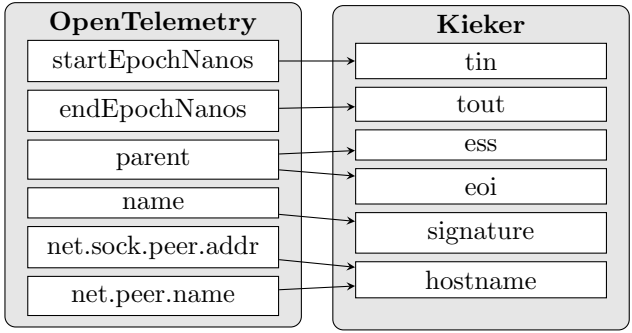

For converting the data, most conversions are just simple renamings: The time stamps and the signature can just be renamed. The spans name is not necessarily a method call, it can also be for example an HTTP call; however, using the name as signature is the best mapping that can be used here. Furthermore, the hostname can be created from a combination of attributes defined in the OpenTelemetry semantic conventions.
Figure~\ref{fig:mapping} represents the mapping of fields.

\paragraph{Control Flow}

For the control flow, OpenTelemetry and Kieker contained not fully compatible concepts: OpenTelemetry represents asynchronous traces, where every span is part of a parent span. One parent span might have multiple child spans taking place at the same time, and the child span that has been started first might end after the child span that has been started second. 

In contrast, Kieker represents synchronous traces, where only one call can take place at the same time. If parallel processing happens, this is represented by separate traces. Internally, Kieker stores an execution order index (\lstinline'eoi') and the execution stack size (\lstinline'ess') of each invocation, which makes it fast to persist the current tracing situation within the tracing agent. The \lstinline'ess' is only allowed to increase by 1, which happens if a method calls a child method.

While OpenTelemetry represents the control flow using a \lstinline'parent' reference, Kieker stores the \lstinline'ess' and \lstinline'eoi' of each invocation. OpenTelemetry's implementation thereby supports asynchronous calls within their traces, Kieker's implementation does not make it necessary to have references in serialized data, which reduces the overhead during tracing.

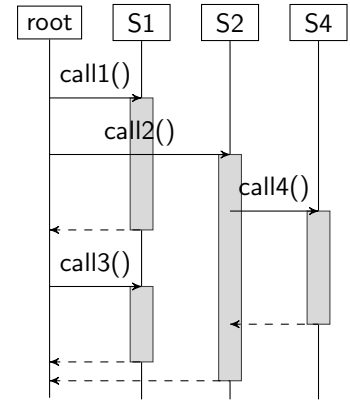
\begin{wrapfigure}{r}{4.4cm}
  \begin{tikzpicture}[>=stealth', font=\sffamily, node distance=4mm and 4mm]

\node[draw, minimum width=0.75cm] (root) {root};
\node[draw, minimum width=0.75cm, right=of root] (S1) {S1};
\node[draw, minimum width=0.75cm, right=of S1] (S2) {S2};
\node[draw, minimum width=0.75cm, right=of S2] (S4) {S4};

\foreach \x in {root, S1, S2, S4} {
  \draw (\x.south) -- ++(0,-4.75) coordinate (\x bottom);
}

\coordinate (t1) at ($(root) + (0,-1)$);
\coordinate (t2) at ($(root) + (0,-1.75)$);
\coordinate (t3) at ($(root) + (0,-2.5)$);
\coordinate (t4) at ($(root) + (0,-2.75)$);
\coordinate (t5) at ($(root) + (0,-3.5)$);
\coordinate (t6) at ($(root) + (0,-4.0)$);
\coordinate (t7) at ($(root) + (0,-4.5)$);
\coordinate (t8) at ($(root) + (0,-4.75)$);

\fill[gray!30] ($(t1 -| S1) + (-0.15cm,0)$) rectangle ++(0.3,-1.75);
\draw[gray!50!black] ($(t1 -| S1) + (-0.15cm,0)$) rectangle ++(0.3,-1.75);

\fill[gray!30] ($(t2 -| S2) + (-0.15cm,0)$) rectangle ++(0.3,-3.0);
\draw[gray!50!black] ($(t2 -| S2) + (-0.15cm,0)$) rectangle ++(0.3,-3.0);

\fill[gray!30] ($(t3 -| S4) + (-0.15cm,0)$) rectangle ++(0.3,-1.5);
\draw[gray!50!black] ($(t3 -| S4) + (-0.15cm,0)$) rectangle ++(0.3,-1.5);

\fill[gray!30] ($(t5 -| S1) + (-0.15cm,0)$) rectangle ++(0.3,-1.0);
\draw[gray!50!black] ($(t5 -| S1) + (-0.15cm,0)$) rectangle ++(0.3,-1.0);

\draw[->] (t1 -| root) -- node[above] {call1()} (t1 -| S1);
\draw[<-, dashed] (t4 -| root) -- node[below] { } (t4 -| S1);

\draw[->] (t2 -| root) -- node[above] {call2()} (t2 -| S2);
\draw[->] (t3 -| S2) -- node[above] {call4()} (t3 -| S4);
\draw[<-, dashed] (t6 -| S2) -- node[below] { } (t6 -| S4);

\draw[->] (t5 -| root) -- node[above] {call3()} (t5 -| S1);
\draw[<-, dashed] (t7 -| root) -- node[below] { } (t7 -| S1);
\draw[<-, dashed] (t8 -| root) -- node[below] { } (t8 -| S2);

\end{tikzpicture}
  \caption{Unrepresentable Parallel Trace}
  \label{fig:asynchronous}
  \vspace{-0.5cm}
\end{wrapfigure}

Since Kieker assumes that traces are sequential, it checks whether parents can be assigned correctly. This process breaks in situations like the one depicted in Figure~\ref{fig:asynchronous}: Here, a root method calls \lstinline'call1' and \lstinline'call2' asynchronously. \lstinline'call1' returns, but \lstinline'call2' still goes on. Later, the root span calls \lstinline'call3', and \lstinline'call2' calls \lstinline'call4'. The spans are ordered by their starting timestamp. Kieker would usually try to assign \lstinline'call4' to a parent, but since \lstinline'call3' already started, there is a gap in the \lstinline'ess' by two, which is a sign of an inconsistent trace. 

To overcome this problems, there are four main solutions:
\begin{inparaenum}[(1)]
  \item Linearize the traces: The calls could be located next to their caller, regardless of when they happen. While this would remove the issue, it would require storing all spans as long as there could be a potential child span arriving, which would increase the resource usage of the trace receiver heavily.
  \item Directly convert the traces: Instead of storing traces as Kieker records into the monitoring log and loading them, OpenTelemetry traces could be transformed into Kieker's \lstinline'ExecutionTrace' directly. This would break the basic split of Kieker's monitoring and analysis components.
  \item Create an additional record that represents asynchronous spans, and can be persisted, loaded and analysed with separate Kieker implementations.
  \item Mark traces as asynchronous: Mark a trace as asynchronous during execution and change the parent assignment if this flag is activated.
\end{inparaenum}

Solution (1) is unacceptable due to the resource usage and solution (2) is unacceptable since it would make repeated analyses impossible. Solution (3) is possible, but would require rewriting big parts of Kieker's analysis pipelines. Therefore, for now, we decied to go with option (4). This requires specifying \lstinline'--asynchronousTrace' to \lstinline'kieker-trace-analysis' when an OpenTelemetry trace is loaded into it. 

\begin{figure*}[hbt]
  \includegraphics[width=\linewidth]{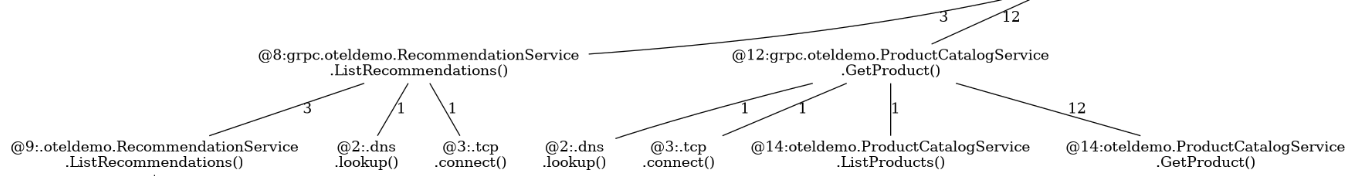}
  \vspace{-0.75cm}
  \caption{Part of Call Tree}
  \label{fig:partOfCallTree}
  \vspace{-0.5cm}
\end{figure*}

This flag is necessary to process all asynchronous traces within the \lstinline'kieker-trace-analysis'. For strictly synchronous traces, e.g., the traces from MooBench, this is not necessary. For traces from microservice applications like the TeaStore \cite{Eismann2018}, the Astronomy shop or the T2 Shop \cite{speth2022saga}, it will always be necessary to set the \lstinline'--asynchronousTrace' flag.

\section{Case Study: Visualizing Astronomy Shop Traces}

The astronomy shop is the default demo application for OpenTelemetry-related tools.\footnote{\url{https://github.com/open-telemetry/opentelemetry-demo}} It consists of 14 services written in 11 different languages, which makes it challenging to instrument them and to establish connections among the agents. Kieker is not able to instrument .NET and TypeScript, therefore, it could not be used to instrument the astronomy shop; however, OpenTelemetry's instrumentation is able to do so.
To showcase our data transformation, we started the astronomy shop with instrumentation and activated our Kieker-otel-transformer. Afterwards, we visualized the obtained trace data using Kieker's trace analysis tool. Figure~\ref{fig:partOfCallTree} shows a part of the created call tree, which visualized the calls from two different services, the product and the recommendation service.

\section{Related Work}

Various works exist that examine the transformation of OpenTelemetry traces. Weber et al. \cite{weber2025integration} examine the interoperability of OpenTelemetry and Palladio. In order to continuously predict the performance, they generate Palladio models based on OpenTelemetry. TraceZip is an approach for storing OpenTelemetry data in a compressed format \cite{chen2025tracezip}. Thereby, they are able to reduce the memory requirement within a TrainTicket-based benchmark by up to 33.8\,\%. Additionally, the throughput is increased, at the cost of increased CPU utilization due to compression. These works  process OpenTelemetry traces, but none of them is able to execute the Kieker analysis with OpenTelemetry traces.
Additionaly, there is research on model transformation in general. Groner et al. \cite{groner2020exploratory} research the view of developers on data model transformations. They find that more than half of the participants already tried to improve the performance of their transformations. While we focused on implementing a prototype in this work, examining the performance of different implementations of the transformation would be valuable future work.

\section{Summary and Outlook}

In this work, we discussed how to achieve the ability to use OpenTelemetry data in the Kieker analysis pipeline. The main issue is the different purpose of the data formats: While Kieker represents sequential traces, OpenTelemetry represents asynchronous traces. We overcome this by marking traces as asynchronous. Thereby, we made it possible to analyse the OpenTelemetry tracing data from the Astronomy Shop using Kieker.
This was the second step of our efforts to make Kieker and OpenTelemetry interoperable. The final step is to support the OpenTelemetry data format fully within Kieker, which requires re-writing parts of the analysis pipeline. Another possible work could be the combination of traces: To have low overhead, applications could be traced inside with Kieker and outside of the application, OpenTelemetry could be used to trace the service calls. By this approach, a combination of Kieker's low-overhead and OpenTelemetry's widespread applicability could be achieved.

\printbibliography

@Book{Blanco2023,
  author = 	 {Daniel Gomez Blanco},
  title = 	 {Practical OpenTelemetry: Adopting Open Observability Standards Across Your Organization},
  publisher = 	 {APress},
  year = 	 {2023},
  doi={10.1007/978-1-4842-9075-0}
}

@inproceedings{weber2025integration,
  title={Integration of performability-model extraction and performability prediction in continuous integration/continuous delivery},
  author={Weber, Sebastian and Weber, Thomas and Hen{\ss}, J{\"o}rg},
  booktitle={SSP 2024},
  journal={Softwaretechnik-Trends},
  Xvolume=45,
  Xnumber=1,
  note = {{\sc pid:} {\tt\href{https://dl.gi.de/handle/20.500.12116/46177}{20.500.12116/46177}}},
  pages={29--31},
  year={2025}
}

@article{chen2025tracezip,
  title={Tracezip: Efficient Distributed Tracing via Trace Compression},
  author={Chen, Zhuangbin and Pu, Junsong and Zheng, Zibin},
  journal={ISSTA},
  pages={411--433},
  year={2025},
  doi={10.1145/3728888}
}

@inproceedings{groner2020exploratory,
  title={An exploratory study on performance engineering in model transformations},
  author={Groner, Raffaela and Beaucamp, Luis and Tichy, Matthias and Becker, Steffen},
  booktitle={ICMDELS},
  pages={308--319},
  year={2020},
  doi = {10.1145/3365438.3410950}
}

@inproceedings{yang2025kieker,
  author = {Yang, Shinhyung and Reichelt, David Georg and Jung, Reiner and Hansson, Marcel and Hasselbring, Wilhelm},
  title = {The Kieker Observability Framework Version 2},
  year = {2025},
  doi = {10.1145/3680256.3721972},
  booktitle = {Companion of the 16th ACM/SPEC International Conference on Performance Engineering},
  pages = {11–-15},
}

@inproceedings{Eismann2018,
  title = {TeaStore: A Micro-Service Reference Application for Cloud Researchers},
  DOI = {10.1109/ucc-companion.2018.00021},
  booktitle = {2018 IEEE/ACM International Conference on Utility and Cloud Computing Companion},
  publisher = {IEEE},
  author = {Eismann,  Simon and Kistowski,  Joakim and Grohmann,  Johannes and Bauer,  Andre and Schmitt,  Norbert and Herbst,  Nikolas and Kounev,  Samuel},
  year = {2018},
  month = dec 
}

@inproceedings{speth2022saga,
  title={A saga pattern microservice reference architecture for an elastic SLO violation analysis},
  author={Speth, Sandro and Stie{\ss}, Sarah and Becker, Steffen},
  booktitle={2022 IEEE 19th ICSA-C},
  pages={116--119},
  year={2022},
  organization={IEEE}
}

@inproceedings{reichelt2025kiekerToOtel,
  author       = {David Georg Reichelt and
                  Malte Hansen and
                  Shinhyung Yang and
                  Wilhelm Hasselbring},
  title        = {Interoperability From Kieker to OpenTelemetry: Demonstrated as Export
                  to ExplorViz},
  booktitle={SSP 2024},
  journal      = {Softwaretechnik-Trends},
  Xvolume       = {45},
  Xnumber       = {1},
  pages        = {20--22},
  year         = {2025},
  note          = {{\sc pid:} {\tt\href{https://dl.gi.de/handle/20.500.12116/46200}{20.500.12116/46200}}},
  bibsource    = {dblp computer science bibliography, https://dblp.org}
}

@InProceedings{reichelt2021overhead,
  title={Overhead Comparison of OpenTelemetry, inspectIT and Kieker},
  author={Reichelt, David Georg and K{\"u}hne, Stefan and Hasselbring, Wilhelm},
  booktitle={SSP 2021},
  year={2021},
  url={https://ceur-ws.org/Vol-3043/}
}

\end{document}